\documentclass[10pt]{article}

\usepackage{amsmath}
\usepackage{amssymb}  
\usepackage[T1]{fontenc}
\usepackage{pifont}
\usepackage{pstricks}
\usepackage{amsfonts}
\usepackage{graphicx}
\usepackage{amsmath}
\usepackage{authblk}
\usepackage{pstricks} 
\usepackage{pstricks-add}
\usepackage{caption}
\usepackage{placeins}
\usepackage{pstricks}
\usepackage{pdfpages}
\usepackage{framed}
\usepackage{cancel}
\usepackage{bm}
\usepackage{epstopdf}
\def\x2dot{\mathop{x}\limits}
\def\y2dot{\mathop{y}\limits}

\def\bfy2dot{\mathop{\bf y}\limits}
\def\z2dot{\mathop{z}\limits}

\def\csi2dot{\mathop{\xi}\limits}
\def\et2dot{\mathop{\eta}\limits}
\def\bet2dot{\mathop{\beta}\limits}
\def\t2dot{\mathop{\theta}\limits}
\def\s2dot{\mathop{\sigma}\limits}
\def\d2dot{\mathop{\delta}\limits}
\def\q2dot{\mathop{q}\limits}
\def\l2dot{\mathop{\lambda}\limits}
\def\ps2dot{\mathop{{\cal E}}\limits}
\def\tet2dot{\mathop{\theta}\limits}
\def\bfx2dot{\mathop{\bf x}\limits}
\def\bfy2dot{\mathop{\bf y}\limits}
\def\bfq2dot{\mathop{\bf q}\limits}
\def\bbfq2dot{\mathop{\bar {\bf q}}\limits}
\def\w2{\mathop{W}\limits}
\def\xgrande2dot{\mathop{\bf X}\limits}
\def\p02dot{\mathop{P}\limits}
\def\a2dot{\mathop{A}\limits}

\newtheorem{prop}{Proposition}
\newtheorem{cor}{Corollary}

\newtheorem{rem}{Remark}
\newtheorem{exe}{Example}

\title{The energy equation for nonholonomic systems with nonlinear kinematic restrictions: a special category of constraints}
\author{F.~Talamucci}
\affil{{\it DIMAI, Dipartimento di Matematica e Informatica ``Ulisse Dini''},\\
{\it	Universit\`a degli Studi di Firenze, Italy}\\
{\it	e-mail: federico.talamucci@unifi.it}}
\date{}

\begin{document}
	\bibliographystyle{plain}
	
	\setcounter{equation}{0}

	\maketitle
	
	\vspace{.5truecm}
	
	
	37J60: Nonholonomic dynamical systems 
	70F25: Nonholonomic systems
	70H03: Lagrange's equations
	70G10  	Generalized coordinates; event, impulse-energy, configuration, state, or phase space for problems in mechanics

	\vspace{.5truecm}
	
	
	\vspace{.5truecm}
	

\begin{abstract}

	\noindent
	The main topic of this work concerns the formulation of the equations of motion and the consequent energy balance that they imply for this type of systems, 
	In particular, the analytical development that we will carry out on the equations of motion has as its objective the energy balance of the system. the delicate question of defining the displacements admitted by the system leads, as we shall see, to a non-univocal definition of the energy of the system, which finds coherence and unity for a particular class of nonholonomic constraints.
	
\end{abstract}

\section{Introduction}

\noindent
The study of mechanical systems subject to kinematic constraints has undoubtedly registered a growing interest, starting from the historical works that laid the foundations of the theory, among which we mention
\cite{appell1}, \cite{capligin1}, \cite{hamel}.
A text of fundamental importance which takes stock of the situation up to that moment and which does the groundwork for many future developments is undoubtedly \cite{neimark}.

\noindent
We can state that the treatment of non-holonomic systems is still today at an incomplete stage in which there is no general theory, comparable to that of holonomic systems, nor a satisfactory understanding of fundamental questions such as the link between symmetries and conservation laws or the extension of Noether's theorem.

\noindent
Truthfully, the category of nonholonomic constraints is wide: the definition and characterization of holonomic and nonholonomic constraints appears in numerous texts on mechanics and
kinematics, based at times on non-unique choices and criteria.
Moreover, in textbooks the treatment of the mechanics of nonholonomic systems is often
reduced to the study of particular systems, developed with ad hoc techniques.
In this regard it should be noted that since the first examples and models studied (for example \cite{appell2}, \cite{voronec2}), typically rigid bodies that slip or roll without sliding on surfaces, find the ideal treatment in the use of the so-called pseudo-coordinates, selected in a targeted way for the problem (\cite{gant} \cite{bloch1}), starting from the theoretical foundations exposed in \cite{hamel2}.
					
\noindent
It must be said that there is a remarkable conceptual and methodological difference between the case of linear kinematic constraints and that of non-linear constraints: if on one hand the first type of constraints  
can be treated convincingly as a plain extension of the holonomic case, 
on the other the case of nonlinear conditions on the kinetic variables does not seem to have a definitive theoretical arrangement.

\noindent
In both cases of linear and nonlinear nonholonomic problems we can broadly outline more than one approach to nonholonomic systems, possibly proposing a mixed way:
\begin{itemize}

\item{} analytical methods, following and extending the Lagrangian formalism which is familiar and consolidated in the case of geometric constraints (among others \cite{zek2}, \cite{zek3}, \cite{cameron}, \cite{li}), 

\item{} an approach via the formalism of Lagrangian multipliers, where the derivatives with respect to the kinetic variables of the constraint equations replace those with respect to the geometric coordinates of the holonomic case (\cite{zeken}, \cite{ranada}, \cite{papa1}, \cite{lurie}),

\item{} methods based on a geometrical support, extending the holonomic formulation via differentiable manifolds to nonholonomic systems, by involving more complex aspects of the theory,as the jet manifold approach (\cite{massa}, \cite{leon}, \cite{sale}),

\item{} a variational approach, essentially investigating the possibility of modifying the principle in order to incorporate the case of nonholonomic constraints (\cite{papa2}, \cite{rumy}); the analysis developed in \cite{cron} and in \cite{flan} shows that the problem is delicate and still open.

\end{itemize}

\noindent
The mathematical investigation presented just after substantially favours the first approach. 

\noindent
In the context of what we have introduced, the present work develops through the following selection of the modelling areas and the following steps. 
We will consider nonholonomic constraints which concern (also in a non-linear way) restrictions on the kinematic state (both geometric coordinates and kinetic variables), possibly depending on time explicitly (rheonomic constraints).
To the same extent that holonomic restrictions allow to identify the essential 
positional parameters, we use the kinematic type restrictions to give rise to a subset of independent generalized velocities, in order to write the equations of motion and develop the theory according to the precepts of the Lagrangian formalism. 
This setting based on the selection among kinematic parameters is inspired by the dated but fundalmental work \cite{voronec1}, where the theory is for linear kinematic restrictions (and still present in the linear version in \cite{neimark}), here extended to the general case of nonlinear kinematic constraints.
 
\noindent
As it is known, for holonomic systems an energy-type equation can be deduced from the equations of motion and the concept of energy of the system can be  opportunely introduced.
If we implement the same formal procedure to the nonoholonomic case we will detect some differences if the starting point is the set of equations in the essential parameters or the set of equations with the lagrangian multipliers (where the independent velocities do not need to be selected). The discrepancy disappears for a remarkable class of kinematic constraints, for which even the notion of virtual displacements theorized to infer equations of motion (${\rm {\check C}}$hetaev condition, \cite{cetaev1}, \cite{cetaev2}, \cite{rumycet}) turns out to be natural and in step with the holonomic case.

\subsection{The equations of motion}

\noindent
The model layout consists in a mechanical system whose configurations are uniquely determined by the set of Lagrangian coordinates $q_1, \dots, q_n$. Besides that, we assume that the system is subject to the following restrictions, involving the generalized velocities and possible the tiem $t$:

\begin{equation}
	\label{constr}
	\begin{cases}
		\phi_1(q_1,\dots, q_n, {\dot q}_1, \dots, {\dot q}_n, t)=0\\
		\qquad \qquad \dots  \dots  \dots\\
		\qquad \qquad \dots \dots \dots\\
		\phi_k(q_1,\dots, q_n, {\dot q}_1, \dots, {\dot q}_n, t)=0
	\end{cases}
\end{equation}

\noindent
In order to write the equations of motion
A direct approach consists in writing the Lagrangian equations of motion 
\begin{equation}
	\label{eqlagrfr}
\dfrac{d}{dt}\dfrac{\partial T}{\partial {\dot q}_i}-\dfrac{\partial T}{\partial q_i}={\cal F}^{(i)}+
{\cal R}^{(i)}, \qquad i=1, \dots, n
\end{equation}
where $T(q_1, \dots, q_n, {\dot q}_1, \dots, {\dot q}_n,t)$ is the kinetic energy of the system and ${\cal F}^{(q_i)}$, 
${\cal R}^{(q_i)}$ are the $i$--th lagrangian components of the acting forces and of the constraint forces, respectively. The structure of the constraints can be outlined by means of the Lagrangian multipliers by setting 
\begin{equation}
	\label{constrmult}
	{\cal R}^{(i)}=\sum\limits_{\nu=1}^k \lambda_\nu \dfrac{\partial \phi_\nu}{\partial {\dot q}_i}, \qquad i=1, \dots, n 
\end{equation}
so that the systems of $n+k$ equations (\ref{eqlagrfr}), (\ref{constr}) contain properly the $n+k$ unknown functions $q_i$, $i=1, \dots, n$, $\lambda_\nu$, $nu=1, \dots, k$.

\noindent
Although the generalized velocities ${\dot q}_i$, $i=1, \dots, n$ are not independent, adopting the point of view (\ref{eqlagrfr}) allows to not explicitly declare the virtual displacements and simply considering the constraints of the type (\ref{constrmult}) as smooth.

\noindent
A quite different formal path consists in explicitly declaring the virtual displacements and directly ignoring the presence of the constraint forces: the first step is to use the restrictions (\ref{constr}) in order to construct the set of virtual displacements.

\noindent
The set of conditions (\ref{constr}) can be made explicit if the following condition holds (without losing generality, except  for re-enumerating the variables):

\begin{equation}
	\label{vincind}
	det\,\left(\dfrac{\partial \phi_i}{\partial {\dot q}_{m+j}}\right)_{i,j=1, \dots, k}\not =0, \quad m=n-k
\end{equation}
In this case it is possible to determine the functions
\begin{equation}
	\label{constrexpl}
	\begin{cases}
		{\dot q}_{m+1}=\alpha_1(q_1, \dots, q_n, {\dot q}_1, \dots, {\dot q}_m, t) \\
		\dots  \\
		\dots  \\
		{\dot q}_n=\alpha_k(q_1, \dots, q_n, {\dot q}_1, \dots, {\dot q}_m, t)
	\end{cases}
\end{equation}
with $m=n-k$. The kinetic parameters ${\dot q}_r$, $=1, \dots, m$, play the role of independent velocities that is, fixing a $m$--uple  ${\dot q}_1, \dots, {\dot q}_m$ establishes the entire kinematic state of the system in the position $q_1$, $\dots$, $q_n$. 

\noindent
In the presence of the setting (\ref{constrexpl}), the formulation of the equations of motion can be based on the assumption that the virtual displacements of the system are
\begin{equation}
	\label{virtdispdip}
	\delta q_{m+\nu}=\sum\limits_{r=1}^m \dfrac{\partial \alpha_\nu}{\partial {\dot q}_r}\delta q_r, \qquad \nu=1, \dots, k
\end{equation}
where the virtual variations $\delta q_1$, $\dots$, $\delta q_r$ are independent. 
By assuming (\ref{virtdisp}), the corresponding equations are 
\begin{equation}
	\label{vnl0}
	\dfrac{d}{dt}\dfrac{\partial T}{\partial {\dot q}_r}-\dfrac{\partial T}{\partial q_r}
	+\sum\limits_{\nu=1}^k\dfrac{\partial \alpha_\nu}{\partial {\dot q}_r}
	\left( \dfrac{d}{dt}\dfrac{\partial T}{\partial {\dot q}_{m+\nu}}-
	\dfrac{\partial T}{\partial q_{m+\nu}}\right)
	={\cal F}^{(r)}+\sum\limits_{\nu=1}^k \dfrac{\partial \alpha_\nu}{\partial {\dot q}_r} {\cal F}^{(m+\nu)}, 
	\qquad r=1,\dots, m.
\end{equation}

\begin{rem}
If ${\bf r}_1$, $\dots$, ${\bf r}_N$ are the vectors which locate the points of the system, assumption (\ref{virtdispdip}) corresponds to 

\begin{equation}
	\label{virtdisp}
	\delta {\bf r}_j = \sum\limits_{r=1}^m \dfrac{\partial {\dot {\bf r}}_j}{\partial {\dot q}_r}\delta q_r, \qquad j=1, \dots, N
\end{equation}
Actually, since
\begin{equation}
	\label{velr}
{\dot {\bf r}_j}=\sum\limits_{r=1}^m\dfrac{\partial {\bf r}_j}{\partial q_r}{\dot q}_r +
\sum\limits_{\nu=1}^k \dfrac{\partial {\bf r}_j}{\partial q_{m+\nu}}\alpha_\nu 
+ \dfrac{\partial {\bf r}_j}{\partial t}, \qquad j=1, \dots, N
\end{equation}
we see that 

$$
\dfrac{\partial {\dot {\bf r}}_j}{\partial {\dot q}_r}=
\dfrac{\partial {\bf r}_j}{\partial q_r}+
\sum\limits_{\nu=1}^k \dfrac{\partial {\bf r}_j}{\partial q_{m+\nu}}
\dfrac{\partial \alpha_\nu}{\partial {\dot q}_r}, \qquad j=1, \dots, N, \quad r=1, \dots, m
$$
hence (\ref{virtdisp}) takes the form
\begin{equation}
	\label{virtdispind}
	\delta {\bf r}_j = \sum\limits_{r=1}^m\left( \dfrac{\partial {\bf r}_j}{\partial q_r}+
	\sum\limits_{\nu=1}^k \dfrac{\partial \alpha_\nu}{\partial {\dot q}_r}\dfrac{\partial {\bf r}_j}
	{\partial q_{m+\nu}}\right)\delta q_r, \qquad j=1, \dots, N
\end{equation}
which entails the same information as (\ref{virtdispdip}) for what concerns the virtual displacements. 
In terms of virtual velocities the assumption (\ref{virtdisp}) can be formally written as
\begin{equation}
\label{velvirtr}
{\widehat {\dot {\bf r}}_j}=
\sum\limits_{r=1}^m \dfrac{\partial {\dot {\bf r}}_j}{\partial {\dot q}_r}{\dot q}_r=
\sum\limits_{r=1}^m \dfrac{\partial {\bf r}_j}{\partial q_r}+
\sum\limits_{\nu=1}^k\dfrac{\partial {\bf r}_j}{\partial q_{m+\nu}}
\sum\limits_{r=1}^m \dfrac{\partial \alpha_\nu}{\partial {\dot q}_r}{\dot q}_r, \qquad j=1, \dots, N
\end{equation}
which does not match the set of velocities consistent with the instantaneous configuration of the system (i.~e.~at a blocked time $t$), obtained by removing the term $\frac{\partial {\bf r}_j}{\partial t}$ in (\ref{velr}).
 \end{rem}

\noindent
Coming back to the equations (\ref{vnl0}), we point out that they can be appropriately written in terms of the only independent generalized velocities ${\dot q}_r$, $r=1, \dots, m$: indeed, by defining the kinetic energy $T^*$ restricted to the constraints (\ref{constrexpl}) as 

\begin{equation}
	\label{trid}
	\begin{array}{l}
		T^*(q_1,\dots, q_n, {\dot q}_1, \dots, {\dot q}_m, t)\\
		\\
		=T(q_1, \dots, q_n, {\dot q}_1, \dots, {\dot q}_m, \alpha_1(q_1, \dots, q_n, {\dot q}_1,\dots, {\dot q}_n,t), \dots, 
		\alpha_k(q_1, \dots, q_n, {\dot q}_1,\dots, {\dot q}_n,t), t)
	\end{array}
\end{equation}
it can be seen that the equations of motion (\ref{vnl0}) take the form (see \cite{tal} for details)

\begin{equation}
	\label{vnl}
	\dfrac{d}{dt}\dfrac{\partial T^*}{\partial {\dot q}_r}-\dfrac{\partial T^*}{\partial q_r}
	-\sum\limits_{\nu=1}^k\dfrac{\partial T^*}{\partial q_{m+\nu}}\dfrac{\partial \alpha_\nu}{\partial {\dot q_r}}
	-\sum\limits_{\nu=1}^k  B_r^\nu \dfrac{\partial T}{\partial {\dot q}_{m+\nu}}
	={\cal F}^{(q_r)}+\sum\limits_{\nu=1}^k \dfrac{\partial \alpha_\nu}{\partial {\dot q}_r} {\cal F}^{(q_{m+\nu})}, 
	\qquad r=1,\dots, m
\end{equation}
where
\begin{equation}
	\label{b}
	B_r^\nu(q_1,\dots, q_n, {\dot q}_1, \dots, {\dot q}_m,t)= 
	\dfrac{d}{dt}\left( \dfrac{\partial \alpha_\nu}{\partial {\dot q}_r}\right)- 
	\dfrac{\partial \alpha_\nu}{\partial q_r}-\sum\limits_{\mu=1}^k 
	\dfrac{\partial \alpha_\mu}{\partial {\dot q}_r}
	\dfrac{\partial \alpha_\nu}{\partial q_{m+\mu}}, \quad r=1, \dots, m, \;\;\nu=1, \dots, k
\end{equation}
Equations (\ref{vnl}) contain the unknown functions $q_1$, $\dots$, $q_n$ but only the derivatives ${\dot q}_r$, ${\q2dot^{..}}_r$ $r=1, \dots, m$ are present, owing to (\ref{constrexpl}); the variables $({\dot q}_{k+1}, \dots, {\dot q}_n)$ that are present in $\frac{\partial T}{\partial {\dot q}_{m+\nu}}$ of (\ref{vnl}) have to be expressed in terms of $(q_1,\dots, q_n,$ ${\dot q}_1, \dots, {\dot q}_m,t)$ by making use of (\ref{constrexpl}). 
The set of equations (\ref{vnl}) joined to (\ref{constrexpl}) represents a system of $m+k=n$ equations in the $n$ unknown functions $(q_1, \dots, q_n)$. Such a system represents an extension to the nonlinear case of the so-called Voronec equations, developed in \cite{neimark} for the linear constraints 
\begin{equation}
	\label{constrlin}
	\left\{
	\begin{array}{l}
		\sum\limits_{i=1}^n\sigma_{1,i}(q_1,\dots, q_n,t){\dot q}_i+\zeta_1(q_1, \dots, \dots, q_n, t)=0\\
		\qquad \qquad \dots  \dots  \dots\\
		\qquad \qquad \dots \dots \dots\\
		\sum\limits_{i=1}^n\sigma_{k,i}(q_1,\dots, q_n,t){\dot q}_i+\zeta_k(q_1, \dots, \dots, q_n, t)=0
	\end{array}
	\right.
\end{equation}
associated to the explicit form (\ref{constrexpl}), whenever the $k\times n$ matrix $(\sigma)_{\nu,i}$ has full rank $k$, according to (\ref{vincind}),
\begin{equation}
	\label{constrlinexpl}
	\begin{cases}
		{\dot q}_{m+1}=\sum\limits_{r=1}^m \alpha_{1,r}(q_1, \dots, q_n, t){\dot q}_r+\beta_1(q_1, \dots, q_n,t) \\
		\dots  \\
		\dots  \\
		{\dot q}_n=\sum\limits_{r=1}^m \alpha_{k,r}(q_1, \dots, q_n, t){\dot q}_r+\beta_k(q_1, \dots, q_n,t)
	\end{cases}
\end{equation}
for suitable coefficients $\alpha_{nu,r}$ and $\beta_r$, $\nu=1, \dots, k$, $r=1, \dots, m$.
In this simpler context, (\ref{virtdispdip}) is 
$$
\delta q_{m+\nu}=\sum\limits_{r=1}^m \alpha_{\nu,r}(q_1, \dots, q_n,t) \delta q_r, \qquad \nu=1, \dots, k.
$$

\begin{rem}
	The case of merely holonomic constraints corresponds to the absence of the conditions (\ref{constr}) and 
	by deleting all the terms containing the $\alpha_\nu$ functions in (\ref{vnl}) we find the usual Euler-Lagrange equations of motion with $T^*=T$.
\end{rem}

\begin{rem}
	Although the set of equations (\ref{vnl}) appears to depend on the selection of the independent velocities, hence the $\alpha_\nu$ functions, the particular dependence on the latter ones only through the derivatives allows us to write the coefficients in terms of the constraint functions: 
	this occurs using the relations that hold for any index $\nu=1, \dots, k$:
	
	\begin{equation}
		\label{derimpl}
	\begin{array}{ll}
		\dfrac{\partial \phi_\nu}{\partial q_i}+\sum\limits_{\mu=1}^k \dfrac{\partial \phi_\nu}{\partial {\dot q}_{m+\mu}}
		\dfrac{\partial \alpha_\mu}{\partial q_i}=0, & i=1, \dots, n \\
		\\
		\dfrac{\partial \phi_\nu}{\partial {\dot q}_r}+\sum\limits_{\mu=1}^k \dfrac{\partial \phi_\nu}{\partial {\dot q}_{m+\mu}}
		\dfrac{\partial \alpha_\mu}{\partial {\dot q}_r}=0, & r=1, \dots, m.
	\end{array}
		\end{equation}
For a fixed $i=1, \dots, n$ and writing the first line for each $j=1, \dots, k$, we obtain a non--singular linear system of $k$ equations in the $k$ unknowns $\dfrac{\partial \alpha_\nu}{\partial q_i}$, $\nu=1, \dots, k$. 
	The same consideration applies to the second set of conditions, where by fixing $r=1, \dots, m$ and letting assume for $j$ the values $1, \dots, k$ we get a non--singular linear system of $k$ equations for the $k$ unknowns 
	$\dfrac{\partial \alpha_\nu}{\partial {\dot q}_r}$, $\nu=1, \dots, k$. 
\end{rem}
\begin{rem}
A point of contact between the two approaches (\ref{eqlagrfr}) (together with (\ref{constrmult})) and (\ref{vnl}) can be found in the so-called ${\rm {\check C}}$hetaev condition (\cite{cetaev1}, \cite{cetaev2}, \cite{rumycet})

\begin{equation}
	\label{cetaev}
	\sum\limits_{i=1}^n\dfrac{\partial \phi_\nu}{\partial {\dot q}_i}\delta q_i=0, \qquad i=1, \dots, \nu
\end{equation}
($\phi_\nu$ are the constraints functions (\ref{constr})). Obviously, (\ref{cetaev}) implies that the virtual 
work of the constraint forces is null, provided that they are formulated as in (\ref{constrmult}):
$$
\sum\limits_{i=1}^n {\cal R}^{(i)}\delta q_i =
\sum\limits_{\nu=1}^k \sum\limits_{i=1}^n \lambda_\nu \dfrac{\partial \phi_\nu}{\partial {\dot q}_i}\delta q_i=0
$$
On the other hand, assumption (\ref{virtdispind}) provides a set of virtual displacements to (\ref{constrmult}), which makes the constraints ideal: actually
\begin{equation}
	\label{rdqi}
\sum\limits_{i=1}^n {\cal R}^{(i)}\delta q_i =	
\sum\limits_{r=1}^m \left( {\cal R}^{(r)}+ 
\sum\limits_{\nu=1}^k {\cal R}^{(m+\nu)}\dfrac{\partial \alpha_\nu}{\partial {\dot q}_r}\right)\delta q_r =
\left( \dfrac{\partial \phi_\nu}{\partial {\dot q}_r}+
\sum\limits_{\nu=1}^k 
\dfrac{\partial \phi_\nu}{\partial {\dot q}_{m+\nu}}\dfrac{\partial \alpha_\nu}{\partial {\dot q}_r}
\right)\delta q_r=0
\end{equation}
owing to the relations (\ref{derimpl}), second group.
\end{rem}

\subsubsection*{The Lagrangian function}

\noindent
If the generalized forces are such that
\begin{equation}
	\label{potu}
	{\mathcal F}^{(i)}=\dfrac{\partial U}{\partial q_i}(q_1, \dots, q_n,t), \quad i=1, \dots, n\quad 
\end{equation}
for a suitable function $U$, then it is possible to define the Lagrangian function ${\cal L}=T+U$ and 
(\ref{eqlagrfr}) (with assumption (\ref{constrmult})) and (\ref{vnl0}) can be written, respectively, as
\begin{eqnarray}
	\label{nnc1}
	\dfrac{d}{dt}\dfrac{\partial {\cal L}}{\partial {\dot q}_i}-\dfrac{\partial {\cal L}}{\partial q_i}=
	\sum\limits_{\nu=1}^k \lambda_\nu \dfrac{\partial \phi_\nu}{\partial {\dot q}_i}
& & i=1, \dots, n\\
\nonumber
\\
\nonumber
\dfrac{d}{dt}\dfrac{\partial {\cal L}}{\partial {\dot q}_r}-\dfrac{\partial {\cal L}}{\partial q_r}
	+\sum\limits_{\nu=1}^k \dfrac{\partial \alpha_\nu}{\partial {\dot q}_r}
	\left( \dfrac{d}{dt}\dfrac{\partial {\cal L}}{\partial {\dot q}_{m+\nu}}-
	\dfrac{\partial {\cal L}}{\partial q_{m+\nu}}\right)=0
& & r=1,\dots, m
	\end{eqnarray}
In turn, the restricted kinetic energy (\ref{trid}) makes us define
\begin{equation}
	\label{lstar}
	{\cal L}^*(q_1, \dots, q_n, {\dot q}_1,\dots, {\dot q}_m, t)=T^*(q_1, \dots, q_n, {\dot q}_1,\dots, {\dot q}_m, t)+U(q_1, \dots, q_n,t).
\end{equation}
In terms of the function (\ref{lstar}) the equations of motion (\ref{vnl}) become
\begin{equation}
	\label{vnlu}
	\dfrac{d}{dt}\dfrac{\partial {\cal L}^*}{\partial {\dot q}_r}-\dfrac{\partial {\cal L}^*}{\partial q_r}
	-\sum\limits_{\nu=1}^k\dfrac{\partial {\cal L}^*}{\partial q_{m+\nu}}\dfrac{\partial \alpha_\nu}{\partial {\dot q_r}}
	-\sum\limits_{\nu=1}^k  B_r^\nu \dfrac{\partial T}{\partial {\dot q}_{m+\nu}}=0, 
	\qquad r=1,\dots, m
\end{equation}
where $B_r^\nu$ are the same functions (\ref{b}).

\section{The energy equation}

\noindent
The next aim is to deduce from (\ref{vnlu}) an information concerning the energy of the system: to this end, let us briefly recall the familiar case of holonomic systems (HC) (which can be identified with $m=n$ and $T^*=T$ in (\ref{trid})): the lagrangian equations of motion imply the following formula:

\begin{equation}
\label{bilenol}
\dfrac{d}{dt}\dfrac{\partial T}{\partial {\dot q}_i}-\dfrac{\partial T}{\partial q_i}={\mathcal F}^{(i)}, \;i=1, \dots, n\;\;\Rightarrow\;\;
\dfrac{d}{dt}
\left(\sum\limits_{i=1}^n {\dot q}_i \dfrac{\partial T}{\partial {\dot q}_i}-T\right)=
\sum\limits_{i=1}^n	{\mathcal F}^{(q_i)}{\dot q}_i-\dfrac{\partial T}{\partial t}
\end{equation}
Furthermore, if (\ref{potu}) holds then $\sum\limits_{i=1}^n {\dot q}_i{\mathcal F}^{(i)}=\dfrac{dU}{dt}-\dfrac{\partial U}{\partial t}$ and (\ref{bilenol}) is simply, by defining with ${\cal L}=T+U$,
\begin{equation}
	\label{bilenollagr}
	\dfrac{d}{dt}\dfrac{\partial {\cal L}}{\partial {\dot q}_i}-\dfrac{\partial {\cal L}}{\partial q_i}=0\;\;\Rightarrow\;\;
	\dfrac{d}{dt}
	\left(\sum\limits_{i=1}^n {\dot q}_i \dfrac{\partial {\cal L}}{\partial {\dot q}_i}-{\cal L}\right)=
-\dfrac{\partial {\cal L}}{\partial t}
\end{equation}
showing the total variation of the hamiltonian function (in round brackets, associable to the energy of the system) with respect to the possible explicit dependence on time of the Lagrangian function;
synthetically we recall that (\ref{bilenol}) is achieved by multiplying each equation of motion by ${\dot q}_i$, by summing the equations and by opportunely rearranging the terms.

\noindent
We examine now the case of nonlinear nonholonomic systems (NNC). 
For our purposes it is reasonable to consider the case of forces (\ref{potu}), since the presence of forces that cannot be traced back to a potential does not affect the meaning of the results.
The equations of motion to be taken into consideration are therefore either (\ref{nnc1}) or
(\ref{vnlu}).

\noindent 
In order to better delineate the definition of energy in nonholonomic systems, let us start with the setting (\ref{nnc1}), first set of equations,  recurrent in literature (\cite{}, \cite{}). Retracing what was done before for (HC) systems, that is multiplying the $i$--th equation by ${\dot q}_i$ and summing up from $1$ to $n$, after rearranginng the terms one gets
\begin{equation}
	\label{bilenmoltipl}
	\dfrac{d}{dt}\left(
	\sum\limits_{i=1}^n 
	{\dot q}_i\dfrac{\partial {\cal L}}{\partial {\dot q}_i}-{\cal L}
	\right)=-\dfrac{\partial {\cal L}}{\partial t}+
	\sum\limits_{\nu=1}^k \sum\limits_{i=1}^n\lambda_\nu \dfrac{\partial \phi_\nu}{\partial {\dot q}_i} {\dot q}_i.
\end{equation}
With respect to (\ref{bilenollagr}), the additional terms of the double sum provide the contribution of the kinematic constraints, by means of the power $\sum\limits_{i=1}^n {\cal R}^{(i)}{\dot q}_i$ of the of the constraint forces (\ref{constrmult}). We stress that the function ${\cal L}$ without the asterisk indicates that no replacement (\ref{constrexpl}) has been made) and we do not expect the power of the constraint forces to be zero.

\noindent
The main aspect that we are going to examine more deeply is the clarifying role of the explicit functions $\alpha_\nu$, $\nu=1, \dots, k$, in order to examine the energy balance of the system. Let us start with the following

\begin{prop}
The constraint forces (\ref{constrmult}) verify
\begin{equation}
\label{potri}
\sum\limits_{i=1}^n{\cal R}^{(i)} {\dot q}_i = 
\sum\limits_{\nu=1}^k
(\alpha_\nu-{\overline \alpha}_\nu)
\left(\dfrac{d}{dt}\dfrac{\partial {\cal L}}{\partial {\dot q}_{m+\nu}}-\dfrac{\partial {\cal L}}{\partial q_{m+\nu}}\right)
\end{equation}
where $\alpha_\nu$, $\nu=1, \dots, k$ are the functions (\ref{constrexpl}) and 
\begin{equation}
\label{baralpha}
{\overline \alpha}_\nu (q_1, \dots, q_n, {\dot q}_1, \dots, {\dot q}_m,t)=
\sum\limits_{r=1}^m \dfrac{\partial \alpha_\nu}{\partial {\dot q}_r}{\dot q}_r.
\end{equation}
\end{prop}

\noindent
{\bf Proof}. The comparison of (\ref{nnc1}) considered for $i=1, \dots, m$ with the immediately following equations implies 
$$
\sum\limits_{\nu=1}^k \lambda_\nu \dfrac{\partial \phi_\nu}{\partial {\dot q}_r}
=-\sum\limits_{\nu=1}^k \dfrac{\partial \alpha_\nu}{\partial {\dot q}_r}
\left(\dfrac{d}{dt}\dfrac{\partial {\cal L}}{\partial {\dot q}_{m+\nu}}-\dfrac{\partial {\cal L}}{\partial q_{m+\nu}}\right), \qquad r=1, \dots, m
$$
thus 
$$
\sum\limits_{\nu=1}^k
\sum\limits_{r=1}^m  \lambda_\nu\dfrac{\partial \phi_\nu}{\partial {\dot q}_r}{\dot q}_r=
-\sum\limits_{\nu=1}^k \overbrace{\sum\limits_{r=1}^m  
\dfrac{\partial \alpha_\nu}{\partial {\dot q}_r} {\dot q}_r}^{={\overline \alpha}_\nu}
\left(\dfrac{d}{dt}\dfrac{\partial {\cal L}}{\partial {\dot q}_{m+\nu}}-\dfrac{\partial {\cal L}}{\partial q_{m+\nu}}\right).
$$

\noindent
On the other hand, considering $i=m+1, \dots, n$ in (\ref{nnc1}) one easily gets
$$
\sum\limits_{\nu, \mu=1}^k \lambda_\mu \dfrac{\partial \phi_\mu}{\partial {\dot q}_{m+\nu}}
{\dot q}_{m+\nu}=
\sum\limits_{\nu=1}^k\overbrace{{\dot q}_{m+\nu}}^{=\alpha_\nu}
\left(\dfrac{d}{dt}\dfrac{\partial {\cal L}}{\partial {\dot q}_{m+\nu}}-\dfrac{\partial {\cal L}}{\partial q_{m+\nu}}\right).
$$
Putting together the last two formulas we overall obtain
$$
\sum\limits_{\nu=1}^k\sum\limits_{i=1}^n  \lambda_\nu \dfrac{\partial \phi_\nu}{\partial {\dot q}_i}{\dot q}_i=\sum\limits_{\nu=1}^k
(-{\overline \alpha}_\nu+\alpha_\nu)
\left(\dfrac{d}{dt}\dfrac{\partial {\cal L}}{\partial {\dot q}_{m+\nu}}-\dfrac{\partial {\cal L}}{\partial q_{m+\nu}}\right)
$$
that is (\ref{potri}). $\quad \square$

\noindent
The property (\ref{potri}) allows us to write (\ref{bilenmoltipl}) in the form
\begin{equation}
	\label{bilene}
	\dfrac{d}{dt}\left(
	\sum\limits_{i=1}^n 
	{\dot q}_i\dfrac{\partial {\cal L}}{\partial {\dot q}_i}-{\cal L}
	\right)=-\dfrac{\partial {\cal L}}{\partial t}-
\sum\limits_{\nu=1}^k
	({\overline \alpha}_\nu-\alpha_\nu)
	\left(\dfrac{d}{dt}\dfrac{\partial {\cal L}}{\partial {\dot q}_{m+\nu}}-\dfrac{\partial {\cal L}}{\partial q_{m+\nu}}\right)
\end{equation}

\noindent
The latter formula shows that the difference with the holonomic case (\ref{bilenollagr}) lies exclusively in the possible difference between $\alpha_\nu$ and the function ${\overline \alpha}_\nu$ defined in (\ref{baralpha}).

\noindent
It is natural at this point to apply the same technique to the equations of motion (\ref{vnlu}) 
in order to obtain an information analogous to (\ref{bilenmoltipl}) involving this time the restricted function ${\cal L}^*$: to this end we prove the following

\begin{prop}

\noindent
The equations of motion (\ref{vnlu}) imply the relation

\begin{equation}
\label{bilenu}
\dfrac{d}{dt}\left( \sum\limits_{r=1}^m {\dot q}_r \dfrac{\partial {\cal L}^*}{\partial {\dot q}_r}
-{\cal L}^*\right)
=
-\dfrac{\partial {\cal L}^*}{\partial t}+
\sum\limits_{\nu=1}^k ({\overline \alpha}_\nu-\alpha_\nu)\dfrac{\partial {\cal L}^*}{\partial q_{m+\nu}}
+\sum\limits_{\nu=1}^k {\overline B}_\nu \dfrac{\partial T}{\partial {\dot q}_{m+\nu}}
\end{equation}
where ${\overline \alpha}_\nu$ is defined in (\ref{baralpha}) and 
\begin{equation}
	\label{barb}
	{\overline B}_\nu (q_1, \dots, q_n, {\dot q}_1, \dots, {\dot q}_m,t)
	=\sum\limits_{r=1}^m B_r^\nu {\dot q}_r, \qquad \nu=1, \dots, k
\end{equation}
with $B_r^\nu$ tbe same coefficients (\ref{b}).
\end{prop}

\noindent
{\bf Proof}. As before, it is sufficient to multiply each equation (\ref{vnlu}) by ${\dot q}_r$ and sum up with respect to the index $r$: bearing in mind the following formula of differentiation 
\begin{equation}
\label{f}
\frac{d}{dt}{\cal L}^*(q_1, \dots, q_n, {\dot q}_1, \dots, {\dot q}_m,t)=
	\sum\limits_{r=1}^m \left( \frac{\partial {\cal L}^*}{\partial q_r}{\dot q}_r
	+\frac{\partial {\cal L}^*}{\partial {\dot q}_r} {\q2dot^{..}}_r \right)+
	\sum\limits_{\nu=1}^k 
	\frac{\partial {\cal L}^*}{\partial q_{m+\nu}}\alpha_\nu+
\frac{\partial {\cal L}^*}{\partial t}
\end{equation}
we see that 
$$
\sum\limits_{r=1}^m{\dot q}_r	
\left(
\frac{d}{dt}\dfrac{\partial {\cal L}^*}{\partial {\dot q}_r}-\dfrac{\partial {\cal L}^*}{\partial q_r}\right)-
\sum\limits_{\nu=1}^k\dfrac{\partial {\cal L}^*}{\partial q_{m+\nu}}\sum\limits_{r=1}^m \dfrac{\partial \alpha_\nu}{\partial {\dot q_r}}{\dot q}_r-
\sum\limits_{\nu=1}^k \sum\limits_{r=1}^m B_r^\nu {\dot q}_r\dfrac{\partial T}{\partial {\dot q}_{m+\nu}}
$$
$$
=\frac{d}{dt}
\left( \sum\limits_{r=1}^m {\dot q}_r \dfrac{\partial {\cal L}^*}{\partial {\dot q}_r}\right)
-\overbrace{\sum\limits_{r=1}^m \left( \dfrac{\partial {\cal L}^*}{\partial q_r}{\dot q}_r
+\dfrac{\partial {\cal L}^*}{\partial {\dot q}_r} {\q2dot^{..}}_r \right)}^{=
\frac{d{\cal L}^*}{dt}-\frac{\partial {\cal L}^*}{\partial t}
-\sum\limits_{\nu=1}^k \frac{\partial {\cal L}^*}{\partial q_{m+\nu}}\alpha_\nu}
-\sum\limits_{\nu=1}^k\dfrac{\partial {\cal L}^*}{\partial q_{m+\nu}} 
{\overline \alpha}_\nu
-\sum\limits_{\nu=1}^k  {\overline B}_r^\nu \dfrac{\partial T}{\partial {\dot q}_{m+\nu}}
$$
from which (\ref{bilenu}) immediately follows. $\quad \square$

\noindent
The formula (\ref{bilenu}) clearly the  nonholonomic contribution originating from the functions $\alpha_\nu$ and the comparison with the holonomic case (\ref{bilenollagr}) is evident: the additional terms appear only if the system undergoes nonholonomic constraints, in the absence of which ${\cal L}^*={\cal L}$ and the usual energy balance for holonomic systems is recovered.

\begin{rem}
It is worth quoting the following formulation of (\ref{barb})
\begin{equation}
	\label{barb2}
	{\overline B}_\nu (q_1, \dots, q_n, {\dot q}_1, \dots, {\dot q}_m,t)
	=\dfrac{d}{dt}({\overline \alpha}_\nu-\alpha_\nu)-\sum\limits_{\mu=1}^k
	\dfrac{\partial \alpha_\nu}{\partial q_{m+\mu}}
	({\overline \alpha}_\mu -\alpha_\mu)+\dfrac{\partial \alpha_\nu}{\partial t}
\end{equation}
the check of which is based on a calculation analogous to (\ref{f}).
The version (\ref{barb2}) shows more clearly the dependence of the coefficients on the deviation of ${\overline \alpha}$ from $\alpha$ and on the possible mobility of the constraints, marked by the presence of the variable $t$ in the functions (\ref{constr}) (rheonomic constraints).
\end{rem}

\begin{rem}
If the forces of the system do not come from a potential then a procedure analogous to the one developed leads to the equation of energy type

\begin{equation}
	\label{bilen}
	\dfrac{d}{dt}\left( \sum\limits_{r=1}^m {\dot q}_r \dfrac{\partial T^*}{\partial {\dot q}_r}
	-T^*\right)-
	\sum\limits_{\nu=1}^k ({\overline \alpha}_\nu-\alpha_\nu)\dfrac{\partial T^*}{\partial q_{m+\nu}}
	-\sum\limits_{\nu=1}^k {\overline B}_\nu \dfrac{\partial T}{\partial {\dot q}_{m+\nu}}
	=-\dfrac{\partial T^*}{\partial t}
	\sum\limits_{r=1}^m {\mathcal F}^{(r)}{\dot q}_r+\sum\limits_{\nu=1}^k {\mathcal F}^{(m+\nu)} {\overline \alpha}_\nu.
\end{equation}

which replaces (\ref{bilenu}); the functions ${\overline \alpha}$ and ${\overline B}_\nu$ are the same as (\ref{baralpha}) and (\ref{barb}). The mixed case of forces partly attributable to a potential and partly not is clear.
\end{rem}

\subsection{The energy function}

\noindent
Let us introduce now the main question of this analysis. The energy-type information  (\ref{bilenollagr}) is formulated using only the independent variables $q_1, dots, q_n$. On the other hand, the expression for nonholonomic systems (\ref{bilenmoltipl}) comes from equations (\ref{nnc1}) , first line, where the kinetic variables are not independent and it could not be otherwise, since the deduced relationships (\ref{constrexpl}) are ignored.
The result (\ref{bilene}) makes us define the quantity in round brackets in the left of the equality 
as the energy of the system. On the other hand, the same comment can be done for (\ref{bilenu})
in order to define the energy in terms of the restricted function ${\cal L}^*$ and to refer to the quantity in round brackets as the energy of the system.

\noindent
It is therefore worthwhile to examine and compare the two expressions that emerge from (\ref{bilenmoltipl}) and (\ref{bilene}), namely
\begin{equation}
	\label{e}
	{\cal E}(q_1, \dots, q_n, {\dot q}_1, \dots, {\dot q}_n,t)=
	\sum\limits_{i=1}^n {\dot q}_i \dfrac{\partial {\cal L}}{\partial {\dot q}_i}-{\cal L}
\end{equation}
and
\begin{equation}
	\label{estar}
	{\cal E}^*(q_1, \dots, q_n, {\dot q}_1, \dots, {\dot q}_m,t)= \sum\limits_{r=1}^m {\dot q}_r 
	\dfrac{\partial {\cal L}^*}{\partial {\dot q}_r}-{\cal L}^*.
\end{equation} 
in order to deal with the question of appropriately defining the energy of the system and with the existence of possible first integrals, which originate uniquely from the balance equations we have introduced above. 

\noindent
Let us start by pointing out that in general the two functions (\ref{e}) and (\ref{estar}) do not coincide: they can actually be compared once they are expressed in the same variables, i.~e.~setting
\begin{equation}
	\label{ecalc}
	{\cal E}(q_1, \dots, q_n, {\dot q}_1, \dots, {\dot q}_m,t)=\left.\left(
	\sum\limits_{i=1}^n {\dot q}_i \dfrac{\partial {\cal L}}{\partial {\dot q}_i}
-{\cal L}\right)\right\vert_{
		{\dot q}_{m+\nu}=\alpha_\nu,\nu=1,\dots, k}
\end{equation}
that is replacing the dependent kinetic variables by means of (\ref{constrexpl}); we indicate again with ${\cal E}$ the recalculated function for simplicity. The function ${\cal E}^*$ calculated by the restricted function ${\cal L}^*$ does not generally coincide with (\ref{ecalc}), whose calculation requires first the use of ${\cal L}$ and then the restriction (\ref{constrexpl}): let us show in the following standard situation the two calculations.

\begin{exe}
	Assume that the kinetic energy is of the standard form 	
	\begin{equation}
		\label{encin}
		T(q_1, \dots, q_n, {\dot q}_1, \dots, {\dot q}_n,t)=\dfrac{1}{2}
		\sum\limits_{i,j=1}^n g_{i,j}{\dot q}_i {\dot q}_j +\sum\limits_{i=1}^n b_i {\dot q}_i+c
	\end{equation}
	where the coefficients $g_{i,j}(q_1,\dots, q_n, t)$ are the entries of a $n\times n$ positive definite matrix 
	and the functions $b_i(q_1,\dots, q_n,t)$, $c(q_1,\dots, q_n,t)$ appear only in the case where the holonomic constraints  depend on time explicitly. Then (\ref{trid})takes the form 
	\begin{equation}
		\label{tmobile}
		T^*=
		\frac{1}{2}\left( \sum\limits_{r,s=1}^m g_{r,s}{\dot q}_r {\dot q}_s+
		\sum\limits_{\nu, \mu=1}^k g_{m+\nu,m+\mu} \alpha_\nu \alpha_\mu \right)+
		\sum\limits_{r=1}^m \sum\limits_{\nu=1}^k g_{r,m+\nu}{\dot q}_r \alpha_\nu+
		\sum\limits_{r=1}^m b_r {\dot q}_r +\sum\limits_{\nu=1}^k b_{m+\nu}\alpha_\nu+c
	\end{equation}
	and the computation of (\ref{estar}), in the validity of  (\ref{potu}), leads to 
	
	\begin{eqnarray}
		\label{enexpl}
		{\cal E}^*&=&
		\frac{1}{2}\sum\limits_{r,s=1}^m 
		g_{r,s}{\dot q}_r {\dot q}_s
		+\sum\limits_{\nu, \mu=1}^k 
		g_{m+\nu, m+\mu}\alpha_\mu ({\overline \alpha}_\nu - \frac{1}{2}\alpha_\nu)\\
		\nonumber
		&+&\sum\limits_{r=1}^m \sum\limits_{\nu=1}^k g_{r,m+\nu} {\overline \alpha}_\nu {\dot q}_r 
		+\sum\limits_{\nu=1}^k b_{m+\nu} ({\overline \alpha}_\nu- \alpha_\nu)-c-U.
	\end{eqnarray}
	On the contrary, the computation (\ref{e}) provides
	\begin{equation}
		\label{eexpl}
	{\cal E}=
	\frac{1}{2}\sum\limits_{r,s}^m g_{r,s}{\dot q}_r {\dot q}_s+
	\dfrac{1}{2}\sum\limits_{\nu, \mu=1}^k 
	g_{m+\nu, m+\mu} \alpha_\nu \alpha_\mu+
	\sum\limits_{r=1}^m \sum\limits_{\nu=1}^k g_{r,m+\nu} \alpha_\nu {\dot q}_r -c-U
\end{equation}
\end{exe}
The Example suggests that the two expressions coincides if 
${\overline \alpha}_\nu = \alpha_\nu$, $\nu =1,\dots, k$: as a matter of facts, the following property shows the deviation between ({\ref{e}) and ({\ref{estar}).

\begin{prop}
	The functions (\ref{e}) and (\ref{estar}) are related by the formula
	\begin{equation}
		\label{rele}
		{\cal E}^*={\cal E}+
		\sum\limits_{\nu=1}^k({\overline \alpha}_\nu-\alpha_\nu)\dfrac{\partial {\cal L}}{\partial {\dot q}_{m+\nu}}.
	\end{equation}
\end{prop}

\noindent
{\bf Proof}.
We write
$$
\sum\limits_{i=1}^n {\dot q}_i \dfrac{\partial {\cal L}}{\partial {\dot q}_i}=
\sum\limits_{r=1}^m {\dot q}_r \dfrac{\partial {\cal L}}{\partial {\dot q}_m}+
\sum\limits_{\nu=1}^k \dfrac{\partial {\cal L}}{\partial {\dot q}_{m+\nu}}\alpha_\nu.
$$
On the other hand, the chain differentiation 
$$
\dfrac{\partial {\cal L}^*}{\partial {\dot q}_r}=\dfrac{\partial {\cal L}}{\partial {\dot q}_r}+
\sum\limits_{\nu=1}^k \dfrac{\partial {\cal L}}{\partial {\dot q}_{m+\nu}}
\dfrac{\partial \alpha_\nu}{\partial {\dot q}_r}\qquad \textrm{for any}\;\;r=1, \dots, m
$$
entails, by multiplying each relation by ${\dot q}_r$ and summing up, 
$$
\sum\limits_{r=1}^m {\dot q}_r\dfrac{\partial {\cal L}^*}{\partial {\dot q}_r}=
\sum\limits_{r=1}^m {\dot q}_r\dfrac{\partial {\cal L}}{\partial {\dot q}_r}+
\sum\limits_{\nu=1}^k\dfrac{\partial {\cal L}}{\partial {\dot q}_{m+\nu}}
\sum\limits_{r=1}^m\dfrac{\partial \alpha_\nu}{\partial {\dot q}_r}{\dot q}_r=
\sum\limits_{r=1}^m {\dot q}_r\dfrac{\partial {\cal L}}{\partial {\dot q}_r}+
\sum\limits_{\nu=1}^k\dfrac{\partial {\cal L}}{\partial {\dot q}_{m+\nu}}
{\overline \alpha}_\nu
$$
owing to definition (\ref{baralpha}). Hence
$$
{\cal E}^*=\sum\limits_{r=1}^m {\dot q}_r\dfrac{\partial {\cal L}}{\partial {\dot q}_r}+
\sum\limits_{\nu=1}^k\dfrac{\partial {\cal L}}{\partial {\dot q}_{m+\nu}}{\overline \alpha}_\nu
-{\cal L}^*, \qquad
{\cal E}=\sum\limits_{r=1}^m {\dot q}_r \dfrac{\partial {\cal L}}{\partial {\dot q}_r}+
\sum\limits_{\nu=1}^k \dfrac{\partial {\cal L}}{\partial {\dot q}_{m+\nu}}\alpha_\nu
-\left.{\cal L}\right\vert_{{\dot q}_{m+\nu}=\alpha_\nu,\nu=1,\dots, k}.
$$
Since 	
$\left.{\cal L}^*(q_1, \dots, q_m, {\dot q}_1, \dots, {\dot q}_n, t)={\cal L}
(q_1, \dots, q_n, {\dot q}_1, \dots, {\dot q}_n, t)\right\vert_{{\dot q}_{m+\nu}=\alpha_\nu,\nu=1,\dots, k}$,
we finally deduce (\ref{rele}). $\quad \square$

\begin{cor}
In the non-singularity situation $\dfrac{\partial T}{\partial {\dot q}_{m+\nu}}\not =0$, $\nu=1, \dots, k$, ${\cal E}^*={\cal E}$ if and only if ${\overline \alpha}_\nu=\alpha_\nu$ for any $\nu=1, \dots, k$.
\end{cor}

\begin{rem}
The relation 
$$
\dfrac{d {\cal E}^*}{dt}=\dfrac{d {\cal E}}{dt}
+\dfrac{d}{dt}\left(\sum\limits_{\nu=1}^k
({\overline \alpha}_\nu-\alpha_\nu)\dfrac{\partial {\cal L}}{\partial {\dot q}_{m+\nu}}\right)
$$
which is obtained by differentiating (\ref{rele}) leads back to (\ref{bilenu}) having in mind (\ref{bilena}), and vice versa: the check is based on (\ref{barb2}) and on the rule
$$
\dfrac{\partial {\cal L}^*}{\partial y}=\dfrac{\partial {\cal L}}{\partial y}+
\sum\limits_{\nu=1}^k \dfrac{\partial {\cal L}}{\partial {\dot q}_{m+\nu}}
\dfrac{\partial \alpha_\nu}{\partial y}\qquad \textrm{for}\;\;\textrm{any of the variables}\;\;\;q_{m+1}, \dots, q_{m+k}, t
$$
\end{rem}

\noindent
The attention paid to the dependence of the various formulas on (\ref{baralpha}) shows that the circustance $\alpha_\nu={\overline \alpha}_\nu$, $\nu=1, \dots, k$ is a special case to be this to be studied separately; 
Before doing this we point out some notable situations, regarding either the structure of the Lagrangian function or that of the constraint functions.

\subsubsection*{The stationary case}

\noindent
We identify the stationary case with systems verifying the assumptions (going back to the general case (\ref{constr}))
\begin{equation}
	\label{statio}
	\dfrac{\partial {\cal L}}{\partial t}=0, \qquad \dfrac{\partial \phi_\nu}{\partial t}=0, \quad \nu =1,\dots ,k
\end{equation}
From the latter condition it follows that also the explicit functions $\alpha_\nu$ do not depend on $t$, that is $\dfrac{\partial \alpha_\nu}{\partial t}=0$, hence $\dfrac{\partial {\cal L}^*}{\partial t}=0$, because of  (\ref{lstar}).
The energy equations (\ref{bilene}) and (\ref{bilenu}) take the form 

\begin{equation}
\label{estatio}
\begin{array}{l}
\dfrac{d {\cal E}}{dt} = 
\sum\limits_{\nu=1}^k
(\alpha_\nu-{\overline \alpha}_\nu)
\left(\dfrac{d}{dt}\dfrac{\partial {\cal L}}{\partial {\dot q}_{m+\nu}}-\dfrac{\partial {\cal L}}{\partial q_{m+\nu}}\right),
\\
\\
\dfrac{d {\cal E}^*}{dt}	=
-\sum\limits_{\nu=1}^k ({\overline \alpha}_\nu-\alpha_\nu)\dfrac{\partial {\cal L}^*}{\partial q_{m+\nu}}
+\sum\limits_{\nu=1}^k 	\left( \dfrac{d}{dt}({\overline \alpha}_\nu-\alpha_\nu)-\sum\limits_{\mu=1}^k
\dfrac{\partial \alpha_\nu}{\partial q_{m+\mu}}
({\overline \alpha}_\mu -\alpha_\mu)\right) \dfrac{\partial T}{\partial {\dot q}_{m+\nu}}
\end{array}
\end{equation}

\noindent
As (\ref{bilenollagr}) recalls, for holonomic systems the energy is conserved in the stationary case: the property cannot be extended to nonholomic systems, unless the property ${\bar \alpha}_u=\alpha_\nu$ for each $\nu=1, \dots, k$ holds.

\noindent
In the context of stationary systems  a special case recurring in the early examples of nonholonomic constraints \cite{} \cite{} concerns the additional assumptions 
\begin{equation}
	\begin{array}{ll}
		\label{chaplin}
		\phi_\nu=\phi_\nu(q_1, \dots, q_m, {\dot q}_1, \dots, {\dot q}_m) & \textrm{for each}\; \nu=1,\dots, k\;\textrm{and}\;j=1, \dots,m\\
		\\\
		{\cal L}={\cal L}(q_1, \dots, q_m, {\dot q}_1, \dots, {\dot q}_n), & 
	\end{array}
\end{equation}
that is the lagrangian coordinates $(q_{m+1}, \dots, q_n)$ corresponding to the dependent velocities are absent. Equations (\ref{vnl}) reduce to 
\begin{equation}
	\label{capligineq}
	\dfrac{d}{dt}\dfrac{\partial {\cal L}^*}{\partial {\dot q}_r}-\dfrac{\partial {\cal L}^*}{\partial q_r}
	-\sum\limits_{\nu=1}^k
	\left(\dfrac{d}{dt}\left(\dfrac{\partial \alpha_\nu}{\partial {\dot q}_r}\right)-
	\dfrac{\partial \alpha_\nu}{\partial \dot q_r} \right) 
	\dfrac{\partial T}{\partial {\dot q}_{m+\nu}}=0
	\quad r=1,\dots, m
\end{equation}	
and they extend the linear $\check{\rm C}$aplygin's equations (see (\cite{capligin1}) and \cite{neimark}) to the nonlinear case.
The evident advantage is that (\ref{capligineq}) contains only the unknown functions $q_1$, $\dots$, $q_m$ and it is disentangled from the constraints equations (\ref{constrlinexpl}).

\noindent
In the same context, assumptions (\ref{chaplin}) make (\ref{estatio}) of the simplified form 
\begin{equation}
	\label{estatiochap}
	\begin{array}{ll}
		\dfrac{d {\cal E}}{dt} = 
		\sum\limits_{\nu=1}^k
		(\alpha_\nu-{\overline \alpha}_\nu)
		\dfrac{d}{dt}\dfrac{\partial {\cal L}}{\partial {\dot q}_{m+\nu}},
		&
		\dfrac{d {\cal E}^*}{dt}	=
	\sum\limits_{\nu=1}^k \dfrac{d}{dt}({\overline \alpha}_\nu-\alpha_\nu)
 \dfrac{\partial T}{\partial {\dot q}_{m+\nu}}
	\end{array}
\end{equation}
which points up the contribution of each kinematic constraint in the variation of the energy. 

\noindent
A further category with particular treatment due to the special form of the Lagrangian function
is represented by those systems where
$T=\dfrac{1}{2}\sum\limits_{i=1}^n M^{(i)}{\dot q}_i^2$ 
with $M^{(i)}$ positive constant (for instance mass of the point associated with the $i$--coordinate)
with the corresponding restricted form (\ref{trid})
\begin{equation}
	\label{tstarcart}
	T^*=\dfrac{1}{2}\sum\limits_{r=1}^m M^{(i)}{\dot q}_r^2+\dfrac{1}{2}\sum\limits_{\nu=1}^k M^{(\nu)}\alpha_\nu^2
\end{equation}
where $\alpha_\nu$, $\nu=1, \dots, k$ are the generic functions (\ref{constrexpl}).
It is simple to verify that the equations of motion (\ref{vnl}) are 
\begin{equation}
	\label{vnlquadr}
	M^{(r)}{\q2dot\limits^{..}}_r+\sum\limits_{\nu=1}^k M^{(m+\nu)}\dfrac{\partial \alpha_\nu}
	{\partial {\dot q_r}} \dfrac{d\alpha_\nu}{dt}=
	{\cal F}^{(q_r)}+\sum\limits_{\nu=1}^{k} \dfrac{\partial \alpha_\nu}{\partial {\dot q}_r}{\cal F}^{(q_{m+\nu})}\quad r=1,\dots, m
\end{equation}
with the obvious modification in case that  (\ref{potu}) holds. In the latter case, (\ref{e}) and (\ref{estar}) are

$$
{\cal E}=
\dfrac{1}{2}\sum\limits_{r=1}^m M^{(r)}{\dot q}_r^2+\dfrac{1}{2}\sum\limits_{\nu=1}^k M^{(\nu)}\alpha_\nu^2
-U, \qquad 
{\cal E}^*=\frac{1}{2}\sum\limits_{r=1}^m M^{(r)}{\dot q}_r^2
+\sum\limits_{\nu=1}^k M^{(\nu)}\alpha_\nu \left({\overline \alpha}_\nu- \frac{1}{2}\alpha_\nu \right)-U, 
$$
and an expressive way to write (\ref{bilene}) or (\ref{bilenu}) is 

\begin{equation}
	\label{bilenplain}
	\dfrac{d}{dt}\left(
	\frac{1}{2}\sum\limits_{r=1}^m M^{(r)}{\dot q}_r^2-U \right)
	+\sum\limits_{\nu=1}^k M^{(\nu)}{\overline \alpha}_\nu \dfrac{d\alpha_\nu}{dt}=
	\sum\limits_{\nu=1}^k ({\overline \alpha}_\nu-\alpha_\nu)\dfrac{\partial U}{\partial q_{m+\nu}}-\dfrac{\partial U}{\partial t}.
\end{equation}
The latter form draws attention to the energy contribution of the $\nu$--th constraint on the applied force, absent if $\overline \alpha_\nu=\alpha_\nu$.

\noindent
In the next Example the function ${\overline \alpha}_\nu$  does not coincide with $\alpha_\nu$.

\begin{exe} 
A peculiar instance of (\ref{tstarcart}) concerns a point $P$ moving in the space and constrained to mantain the same length of the velocity:
	\begin{equation}
		\label{pct}
		|{\dot P}|=C(t)
	\end{equation}
	with $C$ nonnegative function of time. The example is analyzed by \cite{virga}, \cite{swac}) for $C$ constant and by \cite{swac}, \cite{krup} in the case $C(t)=1/\sqrt{t}$. 
	In a coordinate system $(q_1, q_2, q_3)=(x,y,z)$  (\ref{pct}) is ${\dot q}_1^2+{\dot q}_2^2+{\dot q}_3^2-C(t)=0$, so $k=1$ $m=2$ and (\ref{constrexpl}) is
	$$
	{\dot q}_3=\alpha_1({\dot q}_1, {\dot q}_2, t)=\pm \sqrt{C^2(t)-{\dot q}_1^2-{\dot q}_2^2}.
	$$
Assuming for the forces the potential $U(q_1, q_2, q_3)$, evidently ${\cal L}=\dfrac{1}{2}M({\dot q}_1^2+{\dot q}_2^2+{\dot q}_3^2)+U$ and (\ref{tstarcart}) is 
	$T^*=\frac{1}{2}MC^2(t)$, hence ${\cal L}^*= T^*+U={\cal L}^*(q_1, q_2, q_3, t)$.  
 the equations of motion (\ref{vnlquadr}) are 
	
	$$
	\left\{
	\begin{array}{l}
		\dfrac{M}{R}\left( (C^2-{\dot q}_2^2){\q2dot^{..}}_1
		+{\dot q}_1
		({\dot q}_2{\q2dot^{..}}_2-C{\dot C})\right)=
		\dfrac{\partial U}{\partial q_1}\mp \dfrac{{\dot q}_1}{\sqrt{R}}\dfrac{\partial U}{\partial q_1}  \\
		\\
		\dfrac{M}{R}\left( (C^2-{\dot q}_1^2){\q2dot^{..}}_2+{\dot q}_2
		({\dot q}_1{\q2dot^{..}}_1-C{\dot C}) \right)=
		\dfrac{\partial U}{\partial q_2}\mp \dfrac{{\dot q}_2}{\sqrt{R}}\dfrac{\partial U}{\partial q_2} 
	\end{array}
	\right.
	$$
	where we set $M=M^{(1)}$ and  $R({\dot q}_1, {\dot q}_2,t)=C^2(t)-{\dot q_1^2-{\dot q}_2^2}$.
	
\noindent	
Since ${\overline \alpha}_1= \mp\dfrac{C^2-R}{\sqrt{R}}$, ${\overline \alpha}_1-\alpha_1=
\mp\dfrac{C^2}{\sqrt{R}}$, ${\overline \alpha}_1 \dfrac{d \alpha_1}{dt}=-\dfrac{(C^2-R){\dot R}}{2R}$, (\ref{bilenplain}) ca be written as

$$
\dfrac{d}{dt}\left( \frac{M}{2}(C^2-R)-U\right) -\dfrac{M}{2}(C^2-R)\dfrac{\dot R}{R}=\mp \dfrac{C^2}{R}\dfrac{\partial U}{\partial q_3}-
\dfrac{\partial U}{\partial t}
$$ 
where $C^2-R={\dot q}_1^2+{\dot q}_2^2$. In the simpler case $U=U(q_1, q_2)$, the expression we have written may help in order to solve the equations of motion or give indications on the existence of possible first integrals of the motion. In particular, for $C>0$ constant (\ref{bilenplain}) reduces to 
$\dfrac{d}{dt}(\frac{M}{2} y-U)-\frac{M}{2}\frac{y}{C^2-y}{\dot y}=0$, with $y={\dot q}_1^2+{\dot q}_2^2$.

\noindent
In a more general context, (\ref{pct}) can be considered part of the following kinematic condition, involving each singular component:
	$$
	a{\dot x}^2+b{\dot y}^2+c{\dot z}^2=C^2(t)
	$$
	The case $a=b$, $c=-1$, $C(t)=0$ is examined in \cite{fas}. 
\end{exe}

\subsubsection*{Linear kinematic constraints}

\noindent
Many examples and concrete applications show linear kinematic constraints. indeed, most of the literature and theory only consider this case. 
In fact, the most common examples of kinematic constraints such as the pure rolling of a disk or a sphere on a plane or a body that slides frictionlessly with a knife edge that slides only longitudinally can be formulated using linear equations in the velocities and have been  the first studied examples (\cite{capligin2}, \cite{appell2}, \cite{voronec2}).

\noindent
In the case of linear constraints (\ref{constrlin}), we see in the explicit expressions (\ref{constrlinexpl})
that
$\alpha_\nu = \sum\limits_{r=1}^m \alpha_{\nu,r}{\dot q}_r+\beta \nu$ for each $\nu=1, \dots, k$ so that the coefficients (\ref{baralpha}) and (\ref{barb}) are
$$
{\overline \alpha}_\nu=\sum\limits_{r=1}^m \alpha_{\nu,r}{\dot q}_r=\alpha_\nu - \beta_\nu, \qquad 
{\overline B}_\nu=\sum\limits_{j=1}^m\left(
\dfrac{\partial \alpha_{\nu,j}}{\partial t}-\dfrac{\partial \beta_\nu}{\partial t}+
\sum\limits_{\mu=1}^k \left(
\dfrac{\partial \alpha_{\nu,j}}{\partial q_{m+\mu}}\beta_\mu
-\dfrac{\partial \beta_\nu}{\partial q_{m+\mu}}\alpha_{\mu,j}\right)\right)
$$
so that (\ref{bilenu}) is (a similar observation may be done regarding (\ref{bilene}))
$$
\dfrac{d{\cal E^*}}{dt}=
-\dfrac{\partial {\cal L}^*}{\partial t}-
\sum\limits_{\nu=1}^k \beta_\nu \dfrac{\partial {\cal L}^*}{\partial q_{m+\nu}}
+\sum\limits_{\nu=1}^k \sum\limits_{j=1}^m\left(
\dfrac{\partial \alpha_{\nu,j}}{\partial t}-\dfrac{\partial \beta_\nu}{\partial t}+
\sum\limits_{\mu=1}^k \left(
\dfrac{\partial \alpha_{\nu,j}}{\partial q_{m+\mu}}\beta_\mu
-\dfrac{\partial \beta_\nu}{\partial q_{m+\mu}}\alpha_{\mu,j}\right)\right) \dfrac{\partial T}{\partial {\dot q}_{m+\nu}}.
$$
The latter equation shows that, except for the possible explicit dependence on $t$ of the terms, the presence of the functions $\zeta_\nu$ in (\ref{constrlin}), hence of $\beta_\nu$ in (\ref{constrlinexpl}), is the only responsible for the variation of (\ref{estar}). In other words, if ${\cal L}$ does not depend explicitly on time and (\ref{constrlin}) are linear homogeneous functions of ${\dot q}_1$, $\dots$, ${\dot q}_n$, then the energy is conserved.

\noindent
A special situation concerns the affine constraints of the form 
\begin{equation}
	\label{constraff}
	\alpha_\nu=\sum\limits_{j=1}^m a_{\nu,j}(q_1,\dots, q_n, t){\dot q}_j+c_\nu(t), \quad \nu=1,\dots, k
\end{equation}
with $c_\nu$ non zero function. 
In that case ${\bar \alpha}_\nu = \alpha_\nu - c_\nu$, so that  
${\overline B}_\nu=\dfrac{\partial {\overline \alpha}_\nu}{\partial t}$ and (\ref{bilenu}) is
\begin{equation}
	\label{bilenaff}
\dfrac{d{\cal E^*}}{dt}=
-\dfrac{\partial {\cal L}^*}{\partial t}	
-c_\nu\dfrac{\partial {\cal L}^*}{\partial q_{m+\nu}}
+\sum\limits_{\nu=1}^k\left( \sum\limits_{j=1}^m \dfrac{\partial \alpha_{\nu,j}}{\partial t}{\dot q}_j+{\dot c}_\nu\right) \dfrac{\partial T}{\partial {\dot q}_{m+\nu}}.
\end{equation}

\begin{exe}
	For a particle in ${\Bbb R}^3$ with mass $M$ and submitted to the constraint $ax {\dot y}+b {\dot x}y +c -{\dot z}=0$, $c\not=0$, we set $(q_1, q_2, q_3)=(x,y,z)$ so that (\ref{constraff}) writes
	${\dot q}_3=a q_1{\dot q}_2+b q_2{\dot q}_1  +c$ (in this case $m=2$ and $k=1$); moreover 
	$${\cal L}^*=\frac{1}{2}M [(1+a^2 q_1^2){\dot q}_1^2+(1+b^2q_2^2){\dot q}_2^2+2ab q_1q_2 {\dot q}_1{\dot q}_2+2c (aq_1{\dot q}_2 + b q_2 {\dot q}_1)]+U(q_1, q_2, q_3).
	$$
	Whenever $U=U(q_1, q_2)$, (\ref{bilenaff}) supplies the conservation of the quantity
	$$
	\frac{1}{2}M({\dot q}_1^2+{\dot q}_2^2)-\frac{1}{2}M (aq_1 {\dot q}_2 +b q_2 {\dot q}_1 +c)(aq_1 {\dot q}_2 +b q_2 {\dot q}_1 -c).
	$$
	A necessary and sufficient condition (in terms of geometrical properties of the constraint manifold) in order that the energy integral exists is discussed and proved in \cite{fas}.

\end{exe}

\begin{rem}
More generally, for a positive integer $p$ the set of conditions
	\begin{equation}
		\label{constraff}
		\sum\limits_{j=1}^n\sigma_{\nu,j}(q_1, \dots, q_n,t){\dot q}_j^p+\zeta_\nu(q_1, \dots, q_n, t)=0, \quad \nu=1, \dots, k
	\end{equation}
defines affine nonholonomic constraints of degree $p$; the linear case $p=1$ corresponds to (\ref{constrlin}). The 
explicit form (\ref{constrexpl}) which generalizes (\ref{constrlinexpl}) is
	\begin{equation}
		\label{constrexplaff}
		{\dot q}_{m+\nu}^p=(\pm 1)^{p+1}\left(\sum\limits_{j=1}^m \alpha_{\nu,j}(q_1, \dots, q_n,t){\dot q}_j^p+\beta_\nu(q_1, \dots, q_n,t)\right)^{1/p}, \qquad \nu=1, \dots, k
	\end{equation}
for suitable coefficients $\alpha_{\nu,j}$ and $\beta_\nu$. We deduce from (\ref{constrexplaff}):
$$
	\dfrac{\partial \alpha_\nu}{\partial {\dot q}_i}=\dfrac{1}{p} \alpha_\nu
	\dfrac{ \alpha_{\nu,i} {\dot q}_i^{p-1}}
	{\sum\limits_{j=1}^m \alpha_{\nu,j}{\dot q}_j^p+\beta_\nu} \qquad i=1, \dots, m, \quad\nu=1, \dots, k
$$
hence
$$
	{\overline \alpha}_\nu= \dfrac{1}{p}\alpha_\nu \dfrac{ \sum\limits_{i=1}^m\alpha_{\nu,i} {\dot q}_i^p}
	{\sum\limits_{j=1}^m \alpha_{\nu,j}{\dot q}_j^p+\beta_\nu} \qquad \nu=1, \dots, k
$$
so that (\ref{baralphaalpha}) is valid if and only if 
	$$
	(1-p)\sum\limits_{j=1}^m \alpha_{\nu,j}{\dot q}_j^p=p\beta_\nu.
	$$
An evident solution is the case of linear homogeneous constraints ($p=1$, $\beta_\nu=0$ for each $\nu=1, \dots, k$).
	
\end{rem}

\section{A relevant class of nonholonomic constraints}

\noindent
As it emerged, it is worth dwelling on the special case 
\begin{equation}
\label{baralphaalpha}
{\overline \alpha}_\nu=\sum\limits_{r=1}^m {\dot q}_i\dfrac{\partial \alpha_\nu}{\partial {\dot q}_r}=\alpha_\nu\quad \textrm{for each} \quad \nu=1,\dots, k 	
\end{equation}
It is evident that (\ref{bilen}) and (\ref{bilenu}) deserve a distinctive treatment whenever (\ref{baralphaalpha}) holds; such an assumption plays the crucial role for the conservation of the energy of the system. According to what observed in (\ref{derimpl}), the condition depends only on the property of the functions $\phi_\nu$, $nu=1, \dots, k$, and not on the choiche of the explicit functions $\alpha_\nu$.

\noindent
We start by remarking that, whenever (\ref{baralphaalpha}) holds, then the inconsistency (asserted in Remark 1) between the virtual displacements and the velocities compatible with a blocked configuration is overcome: indeed, in this case the expression (\ref{velvirtr}) matches up with the velocity (\ref{velr}) where the last term is dropped.
The property (\ref{baralphaalpha} )therefore extends to nonlinear nonholonomic constraints what happens for holonomic systems and for nonholonomic systems with linear homogeneous constraints, that is (\ref{constrlin})
with $\zeta_\nu=0$, $\nu=1, \dots, k$. 
At the same time, (\ref{potri}) shows that the virtual power of the constraints forces is null, as one can see also from (\ref{rdqi}}).

\noindent
Regarding the energy of systems fulfilling (\ref{baralphaalpha}),  
we state the following property, which pertains to (\ref{e}), (\ref{estar}) and the corresponding energy equations.

\begin{prop}
Assume that (\ref{baralphaalpha}) holds for each $\nu=1, \dots, k$. Then $\sum\limits_{i=1}^n 
{\dot q}_i\dfrac{\partial {\cal L}}{\partial {\dot q}_i}=\sum\limits_{r=1}^m {\dot q}_r \dfrac{\partial {\cal L}^*}{\partial {\dot q}_r}$ so that 
${\cal E}^*={\cal E}$ and (\ref{bilene}), (\ref{bilenu}) reduce to 
	
	\begin{equation}
		\label{bilena}
	\dfrac{d{\cal E}}{dt}=-\dfrac{\partial {\cal L}}{\partial t}\;\;\textrm{or}\;\;
	\dfrac{d{\cal E}^*}{dt}=-\dfrac{\partial {\cal L}^*}{\partial t}
		+\sum\limits_{\nu=1}^k \dfrac{\partial \alpha_\nu}{\partial t}\dfrac{\partial T}{\partial {\dot q}_{m+\nu}}=-\dfrac{\partial {\cal L}}{\partial t}.
	\end{equation}

\noindent
In the stationary case (\ref{statio}), the quantity ${\cal E}={\cal E}^*$ is conserved during the motion.
		
\end{prop}
\noindent
The statements are straightforward consequences (\ref{rele}), (\ref{barb2}) (which implies ${\overline B}_\nu= \dfrac{\partial \alpha_\nu}{\partial t}$) and (\ref{estatio}).
	
\noindent
An remarkable consequence of (\ref{bilena}) is the following
\begin{cor}
Let $\alpha_\nu$, $\nu =1 , \dots, k$ be functions satisfying (\ref{baralphaalpha}). Then, 
the quantity ${\cal E}={\cal E}^*$ is a first integral for the motion (\ref{vnl}) if and only if the Lagrangian function ${\cal L}$ does not depend explicitly on time, that is $\dfrac{\partial {\cal L}}{\partial t}=0$. 
\end{cor}

\noindent
If one refers to the expression (\ref{encin}) for $T$, the Corollary requires that $g_{i,j}$, $b_i$ and $c$,  $i,j=1, \dots, n$,  do not depend on $t$ explicitly and the same goes for $U=U(q_1, \dots, q_n)$; the conserved quantity is the one calculated in (\ref{eexpl}):
\begin{equation}
	\label{intjac}
	\frac{1}{2}\sum\limits_{r,s}^m 
	g_{r,s}{\dot q}_r {\dot q}_s
	+\frac{1}{2}\sum\limits_{\nu, \mu=1}^k 
	g_{m+\nu, m+\mu} \alpha_\nu \alpha_\mu+
	\sum\limits_{r=1}^m	\sum\limits_{\nu=1}^k g_{r,m+\nu} {\dot q}_r\alpha_\nu-U-c=T^*_2-U-c
\end{equation}
where $T^*_2$ (whose definition is clear in (\ref{intjac})) collects the quadratic terms. The absence of the terms containing the coefficients $b_i$, $i=1, \dots, n$, is evidently in line with the Jacobi integral for holonomic systems.
In case of linear kinematic constraints (\ref{constrlin}) the quantity (\ref{intjac}) is (see also (\ref{constrlinexpl})) 
\begin{equation}
	\label{intjaclin}
	\sum\limits_{r,s=1}^m 
	\left(\frac{1}{2} g_{r,s}+\frac{1}{2}\sum\limits_{\nu, \mu=1}^k 
	g_{m+\nu, m+\mu} \alpha_{\nu,r} \alpha_{\mu,s}+
	\sum\limits_{\nu=1}^k g_{r,m+\nu} \alpha_{\nu, s}
	\right){\dot q}_r {\dot q}_s-U-c.
\end{equation}

\begin{exe}
Consider a system of two points $P_1$ and $P_2$ of mass $M_1$ and $M_2$ constrained on a plane and whose velocities are perpendicular and the veloocity of $P_2$ is  parallel to the straight line joining $P_1$ with $P_2$ (or, equivalently, the velocity of $P_1$ is perpendicular to the joining line).	
The constraints equations are 
$$
{\dot x}_1{\dot x}_2+{\dot y}_1{\dot y}_2=0, \qquad (y_1-y_2){\dot x}_2-(x_1-x_2){\dot y}_2=0
$$
and the explicit formulation is, setting $q_1=x_1$, $q_2=x_2$, $q_3=y_1$, $q_4=y_2$, 
	
	$$
	{\dot q}_3=-\dfrac{q_1-q_2}{q_3-q_4}{\dot q}_1, \qquad {\dot q}_4=\dfrac{q_3-q_4}{q_1-q_2}{\dot q_2}
	$$
Having in mind (\ref{constrlinexpl}) with $k=2$ and $m=2$, we see that $\alpha_{1,1}=-(q_1-q_2)/(q_3-q_4)$, $\alpha_{1,2}=\alpha_{2,1}=0$, 
$\alpha_{2,2}=(q_3-q_4)/(q_1-q_2)$, $\beta_1=\beta_2=0$ and the function (\ref{trid}) is                    
	$$
	T^*(q_1, q_2, q_3, q_4, {\dot q}_1, {\dot q}_2)=\dfrac{1}{2}M_1 \left(1+\alpha_{1,1}^2\right){\dot q}_1^2+\dfrac{1}{2}\left(1+\alpha_{2,2}^2\right){\dot q}^2_2
	$$
where $M_1$ and $M_2$ are the masses.
Assuming that the two points are connected by a spring exerting the force $-\kappa (P_1-P_2)$ on $P_1$ and the opposite one on $P_2$ ($\kappa$ positive constant) and including also the gravitational force directed in the direction of decreasing $y$, the Lagrangian function is 
	$$
	{\cal L}=\frac{1}{2}M_1({\dot q}_1^2+{\dot q}_3^2)+\frac{1}{2}M_2({\dot q}_2^2+{\dot q}_2^2)-\frac{\kappa}{2}\left((q_1-q_2)^2+(q_3-q_4)^2\right)-g(M_1q_3+M_2q_4)$$ 
	showing $\dfrac{\partial {\cal L}}{\partial t}=0$; the constant of motion (\ref{intjac}) is 
	$$
I(q_1, q_2, q_3, q_4, {\dot q}_1, 
{\dot q}_2)=T^*+\frac{\kappa}{2}\left((q_1-q_2)^2+(q_3-q_4)^2\right)+g(M_1q_3+M_2q_4).
	$$
\end{exe}

\noindent
We see that the circumstance (\ref{baralphaalpha}) brings the energy balance back to that of holonomic systems (\ref{bilenollagr}), provided that the function ${\cal L}$ or ${\cal L}^*$ is proper positioned in (\ref{bilena}).

\begin{exe}
We consider a system of two material points $(P_1, M_1)$ and $(P_2, M_2)$ on a plane whose velocities are both orthogonal to the straight line joining $P_1$ with $P_2$. In the present case it is $m=k=2$: we set $(q_1, q_2, q_3, q_4)=(x_1, x_2, y_1, y_2)$ and write the constraint equations (\ref{constrlinexpl}) as 
	${\dot q}_3=\frac{q_2-q_1}{q_3-q_4}{\dot q}_1$, ${\dot q}_4=\frac{q_2-q_1}{q_3-q_4} {\dot q}_2$.
Assuming that the active forces verify (\ref{potu}), 
	the Lagrangian function is ${\cal L}=\frac{1}{2}M_1(q_1^2+q_3^2)+\frac{1}{2}M_2(q_2^2+q_4^2)+U(q_1, q_2, q_3, q_4)$ not depending on time and with respect to (\ref{intjaclin})
	one has 
	$g_{1,1}=g_{3,3}=M_1$, $g_{2,2}=g_{4,4}=M_2$, $g_{i,j}=0$ for $i\not =j$, 
	$\alpha_{1,1}=\alpha_{2,2}=\frac{q_2-q_1}{q_3-q_4}$, $\alpha_{2,1}=\alpha_{2,2}=0$
so that the conserved quantity is 
	$$
	I(q_1, q_2, q_3, q_4,{\dot q}_1, {\dot q}_2)=
	\frac{1}{2}(m_1{\dot q}_1^2 +m_2{\dot q}_2^2)\left(1+\left(\frac{q_2-q_1}{q_3-q_4}\right)^2\right)
	-U(q_1, q_2, q_3, q_4).
	$$
\end{exe}

\noindent
Let us comment the time dependence of the constraints (\ref{constr}): the linked equalities in
(\ref{bilena}) make us realize that a first integral can be conferred upon the system even though the nonholonomic constraints
are not stationary, that is the functions (\ref{constrexpl}) depend explicitly on $t$ and so in turn $\dfrac{\partial \alpha_\nu}{\partial t}\not =0$. 
Actually, (\ref{bilena}) ensures that the quantity (\ref{e}) is conserved if and only if ${\cal L}$ does not depend on time explicitly, even though the constraints (\ref{constr}) do: an example in this sense can be sketched as follows.

\begin{exe}
In the three-dimensional space a point $Q\equiv (x_Q, y_Q, z_Q)$ is moving according to the given relations $x_Q=x_Q(t)$, $y_Q=y_Q(t)$, $z_Q=z_Q(t)$. A point $P\equiv (x,y,z)$ can move only pursuing $Q$, that means its velocity has at any time the direction of the straight line joining $P$ and $Q$: the condition $({\dot x}, {\dot y}, {\dot z})\wedge (x_Q(t)-x, y_Q(t)-y,z_Q(t)-z) ={\bf 0}$ gives the two independent constraints 
	\begin{equation}
		\label{constrpurs}
		\left\{
		\begin{array}{l}
			(y_Q(t)-y){\dot x}-(x_Q(t)-x){\dot y}=0, \\
			\\
			(z_Q(t)-z){\dot x}-(x_Q(t)-x){\dot z}=0
		\end{array}
		\right.
	\end{equation}

\noindent
The system consists of only the point $P$ and the lagrangian coordinates can be $q_1=x$, $q_2=y$, $q_3=z$; the explicit form (\ref{constrlinexpl}) is, wherever $x\not= x_Q$, 
	\begin{equation}
		\label{constrpursexpl}
		\left\{
		\begin{array}{l}
			{\dot q}_2 =\dfrac{y_Q(t)-q_2}{x_Q(t)-q_1}{\dot q}_1=\alpha_{1,1}(q_1, q_2, t){\dot q}_1, \\
			\\
			{\dot q}_3 =\dfrac{z_Q(t)-q_3}{x_Q(t)-q_1}{\dot q}_1=\alpha_{2,1}(q_1, q_3, t){\dot q}_1
		\end{array}
		\right.
	\end{equation}
		

\noindent	
The kinetic energy of the system (formed only by the point $P$) is 
$T=\frac{1}{2}M({\dot q}_1^2+{\dot q_2}^2+{\dot q}_3^2)$, where $M$ is the mass, and the form (\ref{trid}) is, owing to (\ref{constrpursexpl}),
\begin{equation}
	\label{encinpurs}
	T^*(q_1,q_2, q_3, {\dot q}_1,t)=\dfrac{1}{2}M{\dot q}_1^2 \left(1+
	\dfrac{(q_2-\eta(t))^2+(q_3-\zeta(t))^2}{(q_1-\xi(t))^2} \right).
\end{equation}
It is immediate to check that the constraints (\ref{constrpursexpl} verify the assumption  (\ref{baralphaalpha})). Assuming (without losing generality) that there are no applied forces, we have ${\cal L}=T$ hence $\dfrac{\partial T}{\partial t}=0$ and the quantity (\ref{e}) (or (\ref{estar})) is conserved even though the constraints are mobile, by virtue of (\ref{bilena}). This quantity is equal to 
${\dot q}_1 \frac{\partial T^*}{\partial {\dot q}_1}=T^*$,  hence
$$
{\dot P}^2=
\frac{2}{M}T^*={\dot q}_1^2\left(1+\dfrac{(q_2-\eta)^2+(q_3-\zeta)^2}{(q_1-\xi)^2}\right)$$ 
is constant during the motion, that is the magnitude of the speed does not vary over time. Such an information can support the resolution of the equation of motion (\ref{vnl}) 
\begin{equation}
	\label{eqpurs}
	\left(1+\dfrac{(q_2-\eta)^2+(q_3-\zeta)^2}{(q_1-\xi)^2}\right)	{\q2dot^{..}}_1=
	-\dfrac{{\dot q}_1{\dot \xi}}{q_1-\xi}\dfrac{(q_2-\eta)^2+(q_3-\zeta)^2}{(q_1-\xi)^2}+
	\dfrac{{\dot q}_1}{(q_1-\xi)^2}
	\left({\dot \eta}(q_2-\eta)+{\dot \zeta}(q_3 - \zeta)\right).
\end{equation}
\end{exe}

\noindent
We finally remark that for systems with $T^*$ as in (\ref{tstarcart}), the validity of (\ref{baralphaalpha}) eliminates in (\ref{bilenplain}) the extra-contribution of the applied forces (with potential $U$); at the same time in that case $\sum\limits_{\nu=1}^k M^{(\nu)}{\overline \alpha}_\nu \dfrac{d\alpha_\nu}{dt}=\dfrac{d}{dt}\left( \dfrac{1}{2}\sum\limits_{\nu=1}^k M^{(\nu)}\alpha_\nu^2\right)$, hence (\ref{bilenplain}) assumes the ordinary form
\begin{equation}
	\label{baltsquarealpha}
	\dfrac{d}{dt}\left(
	\frac{1}{2}\sum\limits_{i=1}^m M^{(i)}{\dot q}_i^2+\frac{1}{2} \sum\limits_{\nu=1}^k M^{(\nu)} \alpha^2_\nu-U \right)=\dfrac{d}{dt}(T^*-U)=
	-\dfrac{\partial U}{\partial t}
\end{equation}
analogous to that pertinent to holonomic systems. 

\subsection{Some special categories of kinematic constraints}

\noindent
In order to intercept the categories of constraints that have this particular treatment (\ref{bilena}) it is 
essential, from the mathematical point of view, to specify which are the functions fulfilling condition (\ref{baralphaalpha}): a first simple but considerable first step is the following 

\begin{prop}
The function ${\overline \alpha}_\nu=\sum\limits_{i=1}^m {\dot q}_i\dfrac{\partial \alpha_\nu}{\partial {\dot q}_i}$ coincides with $\alpha_\nu$ if and only if $\alpha_\nu$ is a positive homogeneous function of degree $1$ w.~r.~t.~${\dot q}_1$, $\dots$, ${\dot q}_m$, 
that is 
\begin{equation}
	\label{hom}
\alpha_\nu (q_1, \dots, q_n, \lambda {\dot q}_1, \dots, \lambda {\dot q}_m, t)=
\lambda \alpha_\nu (q_1, \dots, q_n, {\dot q}_1, \dots, {\dot q}_m, t)\quad \textrm{for any}\;\;\lambda>0.
\end{equation}
\end{prop}

\noindent
{\bf Proof}. It is merely the Euler's homogeneous function theorem: the positively homogeneous functions $f(x_1, \dots, x_N)$ of degree are exactly the solutions of the partial differential equation 
$\sum\limits_{i=1}^N x_i\dfrac{\partial f}{\partial x_i}=kf$; in the present case, the variables $x_1, \dots, x_N$ are the kinetic variables ${\dot q}_1$, $\dots$, ${\dot q}_m$ and $f$ is $\alpha_\nu$, for each $\nu=1, \dots, k$. $\quad\square$

\noindent
Shifting now the attention to the assigned functions (\ref{constr}), it is certainly appropriate to ask how the functions $\phi_\nu$ must be so that the explicit expressions verify (\ref{hom}): we can certainly identify at least the following types of constraints:

\begin{description}
\item[(1)] Linear homogeneous kinematic constraints, that is (\ref{constrlin}) with $\zeta_j=0$:
\begin{equation}
	\label{linom}
		\sum\limits_{i=1}^n\sigma_{\nu,i}(q_1,\dots, q_n,t){\dot q}_i=0 \qquad \nu=1, \dots, k
\end{equation}

\item[(2)] Nonlinear kinematic constraints of type
\begin{equation}
	\label{quadrom}
	\sum\limits_{i,j=1}^n a_{i,j}^{(\nu)}(q_1, \dots, q_n,t){\dot q}_i {\dot q}_j=0, \qquad \nu=1, \dots, k
\end{equation}

\noindent
that is homogeneous quadratic functions with respect to the kinetic variables.
\end{description}

\noindent
The check of (\ref{hom}) for the linear case (\ref{linom}) is immediate; as for the case (2), it can be 
seen by means of the implicit function theorem that the functions (\ref{constrexpl}) defined by (\ref{quadrom}) are actually homogeneous functions of degree $1$ with respect tothe variables ${\dot q}_1$, $\dots$, ${\dot q}_r$. 

\noindent
Examples of the typology (\ref{quadrom}) are standard kinematic restrictions which are frequently considered in applications and in the literature (we limit ourselves to systems of only two points for ease of writing):

$$
\begin{array}{ll}
	{\dot x}_1^2+{\dot y}_1^2+{\dot z}_1^2-\left({\dot x}_2^2+{\dot y}_2^2+{\dot z}_2^2\right)=0& \textrm{same magnitude of the velocities} \\
	{\dot x}_1{\dot y}_2-{\dot x}_2{\dot y}_1=0, \quad {\dot x}_1{\dot z}_2-{\dot x}_2{\dot z}_1=0 
& \textrm{parallel velocities} \\
	{\dot x}_1{\dot x}_2+{\dot y}_1{\dot y}_2 + {\dot z}_1 {\dot z}_2 = 0 & \textrm{perpendicular velocities} 
\end{array}
$$

\noindent
The mentioned systems have access to the balance equation (\ref{bilena})
and the absence of $t$ from the Lagrangian function ${\cal L}$ will provide the first integral of motion (\ref{estar}).

\noindent
For the examples listed above the coefficients  $a_{i,j}^{(\nu)}$ in (\ref{quadrom}) are constant: in the following final example, known as a nonholonomic pendulum, the coefficients depend on the Lagrangian coordinates.

\begin{exe}
		Two points $P_1$ and $P_2$ are moving on a
	plane in a way that the straight lines perpendicular to the velocities of the points ${\dot P}_1$, ${\dot P}_2$ intersect in one of the points of a given curve $\gamma$ lying on the plane. 
	
If $(x_i,y_i,0)$ are the coordinates of $P_i$, $i=1,2$, the two straight lines perpendicular to the velocities intersect at the point of coordinates 
	$$
	{\bar x}=
	\dfrac{{\dot y}_1{\dot y}_2(y_1-y_2)-x_2{\dot y}_1 {\dot x}_2+x_1{\dot x}_1{\dot y}_2}
	{{\dot x}_1{\dot y}_2 - {\dot x}_2{\dot y}_1},\;\;
	{\bar y}=\dfrac{{\dot x}_1{\dot x}_2(x_2-x_1)-y_1{\dot y}_1 {\dot x}_2+y_2{\dot x}_1{\dot y}_2}
	{{\dot x}_1{\dot y}_2 - {\dot x}_2{\dot y}_1}
	$$
	so that, if the curve $\gamma$ is the graph of $y=g(x)$, the kinematic constraint is formulated by
	$$
	{\bar y}(x_1, y_1, x_2, y_2, {\dot x}_1, {\dot y}_1, {\dot x}_2, {\dot y}_2)=g({\bar x}(x_1, y_1, x_2, y_2, {\dot x}_1, {\dot y}_1, {\dot x}_2, {\dot y}_2))
	$$
Let us assume for simplicity that $\gamma$ is the straight line $ax+by=0$: the constraint is
\begin{equation}
	\label{nonolpend}
	a\left({\dot y}_1{\dot y}_2(y_1-y_2)-x_2{\dot y}_1 {\dot x}_2+x_1{\dot x}_1{\dot y}_2\right)+
	b\left({\dot x}_1{\dot x}_2(x_2-x_1)-y_1{\dot y}_1 {\dot x}_2+y_2{\dot x}_1{\dot y}_2\right)=0.
\end{equation}
Wih respect to (\ref{constr}), we have $n=4$ and $k=1$ (hence $m=3$): setting $q_1=x_1$, $q_2=y_1$,  $x_2=q_3=x_2$, $q_4=y_2$, the explicit form of (\ref{nonolpend}) is
	\begin{equation}
		\label{nhpexpl}
		{\dot q}_4 = \dfrac{b(q_1-q_3){\dot q}_1+(bq_2+aq_3) {\dot q}_2}
		{(aq_1+bq_4){\dot q}_1+a(q_2-q_4){\dot q}_2}{\dot q}_3 = \alpha_1 (q_1, q_2, q_3, q_4, {\dot q}_1, {\dot q}_2, {\dot q}_3)
	\end{equation}
The constraint (\ref{nonolpend}) is of (\ref{quadrom}) type and we see that the explicit form (\ref{nhpexpl}) is a homogeneous function of degree $1$ with respect to the kinematic independent variables ${\dot q}_1$, ${\dot q}_2$, ${\dot q}_3$.
Assuming that the forces give rise to the potential $U(q_1, q_2, q_3, q_4)$, neither the function 
		$$
		{\cal L}^*=\dfrac{1}{2}M_1 ({\dot q}_1^2+{\dot q}_2^2)+\dfrac{1}{2}M_2 {\dot q}_3^2\left(
		 1+\left(\dfrac{b(q_1-q_3){\dot q}_1+(bq_2+aq_3) {\dot q}_2}
		{(aq_1+bq_4){\dot q}_1+a(q_2-q_4){\dot q}_2}\right)^2\right)+U(q_1, q_2, q_3, q_4)
		$$
nor the constraint (\ref{nhpexpl}) depend explicitly on time $t$. The conserved quantity coming from (\ref{bilena}) is obtained by inverting the sign before $U$ in ${\cal L}^*$.
\end{exe}
	
\section{Conclusion and upcoming research}

\noindent
The central theme of the first part of the work is the condition \ref{cetaev}) that although it dates back to over seventy years ago, it is accepted directly as an axiom (anyway in accordance with physical measurements) rather than having a provenance from the laws of mechanics. 

\noindent
if one follows the Lagrange multiplier procedure, this hypothesis ensures that the constraint forces do not perform virtual work.
If instead we turn to a subset of independent velocities by means of (\ref{constrexpl}, the same hypothesis provides the set of virtual displacements useful for writing the equations of motion (\ref{vnl}).

\noindent
The fundamental difference that brings systems with nonlinear constraints to a more complicated level than holonomic systems and those with linear constraints lies in the fact that for the latter ones the notion of displacement overlaps with that of virtual velocity. On the contrary, the formula (\ref{velvirtr}) does not trace a virtual velocity seeing as the components ${\dot q}_{m+\nu}=\alpha_\nu$, $\nu=1, \dots, k$, are replaced by the functions ${\overline \alpha}_\nu$ defined in (\ref{baralpha}).
Such a discrepancy can be approached to the analysis carried out in \cite{papa2} (even though in a much more elaborate formal context), where the concept of admissible or virtual displacements are explored and discriminated.

\noindent
Inevitably the standard procedure in the Lagrangian formalism of producing energy-type information from the equations of motion, which we dealt with in the second part of the work,
entails also in this case a difference with the holonomic or linear nonholonomic situation for the reason, essentially, that the powers of forces cannot directly emerge in the energy equation (\ref{bilen}) and in the similar ones.

\noindent
It is therefore legitimate to wonder how much the Hamiltonian type function (\ref{estar}) built with only the independent velocities can be assumed as the energy of the system: the discrepancy that we first reported means that this function does not necessarily correspond to the energy (\ref{e}) calculated with all the variables.

\noindent
The care we took in highlighting the dependence of the various formula (firt of all (\ref{rele})) on the differences ${\overline \alpha}_\nu-\alpha_\nu$ leads us to conclude that the hypothesis (\ref{baralphaalpha})is essential and indicative for treating the case of nonlinear nonholonomic constraints as a plain extension of the linear case.
Under the same assumption, the discrimination between ``virtual'' and ``admissible'', to say, does not exist.

\noindent
The conclusion leads to further topics to investigate: which types of constraints (\ref{constr}) make the hypothesis (\ref{baralphaalpha}) hold?
The mathematical conjecture is the following: if all the functions $\phi_nu$ in (\ref{constr}), $\nu=1, \dots, k$, are homogeneous functions (even if of different degree) with respect to the kinetic variables ${\dot q}_1$, $\dots$, ${\dot q}_n$, then the explicit functions (\ref{constrexpl}) are homogeneous functions (with respect the independent velocities) of degree one. 

\noindent
On the other hand, one may suppose that the zero set  defined by (\ref{constr}) can be always realized by means of homogeneous functions (it is known that the same constraints, even kinematic ones, can be realized by different sets of conditions).

\noindent
In conclusion, the topic to be explored  is whether the disentangling condition (\ref{baralpha}) is a prerogative for systems with homogeneous nonholonomic constraints, clearing the way for claiming that only this type of systems can show (under further assumptions) the conservation of energy.

\end{document}